\documentclass[preprint,12pt]{elsarticle}

\usepackage{amssymb}

\usepackage{bussproofs}
\usepackage{listings}
\usepackage{lipsum}
\usepackage{xcolor}
\usepackage{amsmath}
\usepackage{comment}
\usepackage{url}

\newcommand{\marcos}[1]{{#1}}

\let\origendprooftree=\endprooftree
\def\endprooftree{\origendprooftree\vspace{1mm}}

\let\origRightLabel=\RightLabel
\renewcommand{\RightLabel}[1]{\origRightLabel{\small{#1}}}

\journal{Journal of Computer Languages}

\begin{document}

\begin{frontmatter}



\title{A Gradual Type System for Elixir}


\author[inco]{Mauricio Cassola}
\ead{mauricio.cassola@fing.edu.uy}

\author[inco]{Agustín Talagorria}
\ead{agustin.talagorria@fing.edu.uy}

\author[inco]{Alberto Pardo}
\ead{pardo@fing.edu.uy}

\author[inco]{Marcos Viera}
\ead{mviera@fing.edu.uy}

\affiliation[inco]{organization={Instituto de Computacion, Universidad de la Republica},
            city={Montevideo},
            country={Uruguay}}


\begin{abstract}
Elixir is a functional programming language with dynamic typing. We propose a gradual type system that makes it possible to perform type-checking on a significant fragment of the language. An important feature of the type system is that it does not require any syntactic change to Elixir. Type information is provided by means of function signatures which are declared in terms of Elixir typespec directives. The proposed type system is based on subtyping and is backward compatible, as it allows the presence of untyped code fragments. We have implemented a prototype of the type-checker in Elixir itself.
\end{abstract}




\end{frontmatter}



\section{Introduction}

Elixir~\cite{elixir} is a functional general-purpose programming language that runs on the Erlang virtual machine (BEAM)~\cite{beam}.
It focuses on developer productivity, with support for concurrency and agile web programming as its backbone.
Elixir takes advantage of all the benefits offered by the Erlang VM and its ecosystem, while its syntax is influenced by a modern programming language such as Ruby~\cite{ruby}. It also comes with a toolset that simplifies project management, testing, packaging, and documentation building.



Like Erlang, Elixir is a language with dynamic typing. There are tools, such as \emph{Dyalixir}~\cite{dialyxir}, that make it possible to perform some kind of static typing in Elixir. Dyalixir simply connects with a popular Erlang tool, called \emph{Dialyzer}~\cite{dialyzer}, which performs static analysis on BEAM bytecode. It is based on the concept of \emph{success typing}~\cite{LS06} to find type errors in the code. Similar to \emph{soft types} \cite{CF04}, success typing suffers from the disadvantage of being quite permissive, while it is flexible and more adaptable to the dynamic typing style. 
Error messages in Dialixir are not very descriptive of the place where the errors in the original Elixir code are or why they happen, because the analysis is performed on BEAM bytecode.




In this paper we introduce a type system for a fragment of Elixir that makes it possible to perform type checking without losing the essence and flexibility of the language. 
Our approach is inspired by so-called \emph{gradual typing}~ \cite{ST06,ST07,CLP+19}, which is present in extensions of many dynamic languages like  TypeScript, PHP, Typed Racket and Typed Clojure, among others. The proposed type system is based on subtyping and is backward compatible, as it allows the presence of untyped code fragments. 
Our type system follows Castagna et al. \cite{castagna} approach in that it uses an order relation between gradual types (called the \emph{precision relation}) with an associated downcast rule; this is in contrast with most gradual type systems which use a consistency relation on gradual types.

Unlike success typing, where possible type clashes are detected by type inference, in gradual typing the programmer decides which parts of the code have to be statically type checked. 
Type information is essentially provided by means of function signatures which are declared in terms of Elixir's type specification  directive (\texttt{@spec}).
An important feature of our type system is that it does not produce any change to the analysed code. The code parts that cannot be statically type checked because of lack of typing information are then left to be type checked at runtime.


In Section~\ref{sec:elixir} we give a brief description of the Elixir fragment we consider and we introduce the main features of our type system by means of examples. Section~\ref{sec:typesystem} presents the formal type system; first, we introduce a static type system and then we extend it in order to deal with gradual typing. Finally, in Section~\ref{sec:conclusions} we conclude and present some directions for future work.

This is an extended and revised version of the work presented at SBLP 2020 \cite{CTPV20}. In this paper we extended the language features covered by the type system with anonymous functions, variable application, higher-order functions and the pin operator. We have also modified the language of types, having now atoms as singleton types with the \texttt{atom} type as their supertype.
An important change with respect to the SBLP paper is that we have split the unknown and top types into two different types \texttt{any} and \texttt{term}. This made it possible to remove some restrictions on subtyping we had before that were caused by the identification of these two types in one. The separation of these two types is in accordance with most of the literature in gradual typing.

\section{Towards a Typed Elixir}\label{sec:elixir}

The grammar in Figure~\ref{fig:grammar} describes the fragment of Elixir we are considering for typing. It is important to emphasize that we are not making any changes to the language's syntax. Function signatures are provided by using the \texttt{@spec} directive,
using the notation for declaring types 
existing in Elixir.


\begin{figure}
\begin{tabular}{ l c l }
 $prog$ & ::= & $m \; | \;  e$  \\
 $m$ & ::= & $m;m \; | \; \texttt{defmodule } m\!\_name \texttt{ do } bdy  \texttt{ end}$\\
 $bdy$ & ::= & $m \; | \;  e \; | \;  f$ \\
 $e$ & ::= & $l \; | \; x  \; | \; \texttt{\{}e\texttt{,} ...\texttt{,} e\texttt{\}} \; | \; \texttt{[}\ \texttt{]} \; | \; \texttt{[}e \texttt{|} e\texttt{]} \; | \; \texttt{\%\{}k \Rightarrow e\texttt{,} ...\texttt{,} k \Rightarrow e\texttt{\}} \; | \; e\texttt{[}k\texttt{]}$ \\
& $|$ & $e \ \oplus \ e \; | \;  \ominus \ e$ \\
& $|$ & $\texttt{if }  e \texttt{ do } e  \texttt{ else } e \texttt{ end}$\\  
  & $|$ & $\texttt{case } e \texttt{ do } p \rightarrow e \texttt{; }...\texttt{; } p \rightarrow e \texttt{ end}$ \\
  & $|$ & $\texttt{cond do } e \rightarrow e \texttt{; } ... \texttt{; } e \rightarrow e \texttt{ end}$ \\
  & $|$ & $q\ f\!\_name (e\texttt{,} ...\texttt{,} e) \; | \; x  \texttt{.} (e\texttt{,} ...\texttt{,} e) \; | \; \texttt{fn } (p, ..,p) \rightarrow e \texttt{ end}$ \\
  & $|$ & $ p \texttt{ = } e \; | \; e ; e$ \\
 $k$ & ::= & $a \; | \;  b \; | \;  i$ \\
 $l$ & ::= & $k \; | \;  r \; | \;  s$ \\
 $q$ & ::= & $m\!\_name.q \; | \; \epsilon$ \\
 $p$ & ::= & $\_ \; | \; l \; | \; x \; | \; \texttt{\^{}}x \; | \; \texttt{\{}p \texttt{, }...\texttt{, } p\texttt{\}} \; | \; \texttt{[}\ \texttt{]} \; | \; \texttt{[}p \texttt{|} p\texttt{]} \; | \; \texttt{\%\{}k \Rightarrow p \texttt{, } ... \texttt{, } k \Rightarrow p\texttt{\}}$ \\
 $f$ & ::=  & $\texttt{@spec } f\!\_name(t\texttt{,} ...\texttt{,} t) \texttt{ :: } t$ \\
 & $|$ & $\texttt{def} f\!\_name(p \texttt{, } ... \texttt{, } p) \texttt{ do } e \texttt{ end}$ \\
 & $|$ & $f;f$ \\ 
  $t$ & ::= & $\texttt{none} \; | \;   \texttt{term} \; | \;  \texttt{any}$ \\ 
  & $|$ & $\texttt{integer}  \; | \; \texttt{float} \; | \;  \texttt{boolean} \; | \;  \texttt{string} \; | \;  \texttt{atom} \; | \; a$ \\  
  & $|$ & $[t] \; | \; \{t,...,t\} \; | \; \%\{k \Rightarrow t,..., k \Rightarrow t\}$ \\
  & $|$ & $(t,...,t) \rightarrow t$ \\

  \multicolumn{3}{l}{$m\!\_name, x, f\!\_name,  \in Id$} \\
  
\end{tabular}

\hrule
\caption{Grammar of Elixir fragment}
\label{fig:grammar}
\end{figure}


\subsection{Elixir}\label{subsec:elixir}

We start with a brief description of Elixir. A program in Elixir is composed by a sequence of either modules or expressions.
Expressions include literals, variables, tuples, lists, maps, binary and unary operators, conditional expressions,  function applications, anonymous function definitions and matchings.

Literals ($l$) can be integer ($i$) or float ($r$) numbers, strings ($s$), boolean constants ($b$) or atoms ($a$); atoms are constants whose values are their own name, starting with a colon  (e.g. \texttt{:ok}, \texttt{:error}, \texttt{:apple}). 

We write $\oplus$ to denote the binary operators: arithmetic   (\texttt{+, -, *, /}), boolean (\texttt{and, or}), relational (\texttt{<, >, <=, >=, ==, !=, ===, !==}), list append ($++$), list remove ($--$), and string concatenation (\texttt{<>}).
The unary operators ($\ominus$) are substraction (\texttt{-}) and boolean negation (\texttt{not}). 

A sequence of expressions \texttt{e1;...;en} evaluates to the value of its last expression \texttt{en}. Variables can be bound to values through so-called matchings, in order to use them later in the sequence.
\begin{lstlisting}
x = 10 * 9
x + 10                                # 100           
\end{lstlisting}

\noindent
Variables can be rebound multiple times and their scope is limited to the block where they are defined. For example, the following sequence of expressions evaluates to the tuple \texttt{\{1,3\}}:
\begin{lstlisting}
x = 1         
y = if true do x = 2; x + 1 else 4 end 
{x,y}                                 # {1, 3}
\end{lstlisting}

We can construct structured data using tuples, lists and maps. As usual, a tuple \texttt{\{e1,...,en\}} contains a fixed number of elements in a given order. Lists are constructed using cons cells with the syntax \texttt{[head | tail]}; the empty list is written as \texttt{[]}. Maps are collections of key-value pairs.
Some examples:

\begin{lstlisting}
t = {"one",2,:three}                  # tuple
l = [ 9 | [ "hi" | [ true | [] ] ] ]  # list
m = %{:strange => "map", 9 => true}   # map
\end{lstlisting}
Observe that Elixir permits heterogeneous lists and maps with keys of different types.
Map lookup is written using the following notation:
\begin{lstlisting}
m[9]                                  # true
\end{lstlisting}

Structured values (lists, maps and tuples) can be deconstructed by applying pattern matching:
\begin{lstlisting}
[ x | _ ] = l
x + 10                                # 19           
\end{lstlisting}
A map pattern matches a map if it contains the keys and their respective values match.
\begin{lstlisting}
case m do %{:strange => x} -> x end   # "map"

%{9 => b} = m
b                                     # true
\end{lstlisting}

Patterns are non-linear. Therefore, multiple occurrences of a variable in a pattern must be bound to the same value. For instance, the following expression does not match because the variable \texttt{x} is bound to two different values.
\begin{lstlisting}
{x,x} = {2,3}            # matching error
\end{lstlisting}
The \emph{pin operator} \texttt{\^{}x} is used within patterns when one wants to match the current value of the variable \texttt{x}, if instead of binding it to a new value.
For example:
\begin{lstlisting}
x  = 1
^x = 2                   # matching error
\end{lstlisting}

\begin{lstlisting}
x = 1
{^x,y} = {1,2}           # correct matching
\end{lstlisting}

Functions are identified by their name and arity. Thus, different functions with the same name can coexist and partial application is not allowed. Functions are defined with the keyword \texttt{def}, declaring their name, parameters and body. Multiple definitions of a function are allowed. To apply a function, the argument list is enclosed in parentheses.
The names of non-local functions are prefixed with the hierarchy of the module to which they belong.
\begin{lstlisting}
defmodule Base do
    defmodule Math do
        def dec(x) do x - 1 end
    end
end

defmodule Main do
    def fact(0) do 1 end
    def fact(n) do n * fact(Base.Math.dec(n)) end
end
\end{lstlisting}
Notice the use of pattern matching to define function \texttt{fact}.

We can define anonymous functions in expressions, and thus bind them to variables.
\begin{lstlisting}
f = fn (x) -> x + 1 end
\end{lstlisting}
Such functions are applied using the following syntax:
\begin{lstlisting}
f.(8)                                # 9           
\end{lstlisting}
By using anonymous functions we can define higher-order functions.
\begin{lstlisting}
def twice(g,v) do g.(g.(v)) end
\end{lstlisting}
Then, if for example we apply $\texttt{twice(f,2)}$ the result is $4$.

\subsection{Typed Elixir}\label{sec:typed}

Elixir posseses some basic types, such as $\texttt{integer}$, $\texttt{float}$, $\texttt{string}$, $\texttt{boolean}$ and $\texttt{atom}$, which are dynamically checked at runtime.
Type specifications (\texttt{@spec}) have no meaning for the compiler; they are used to document code or by tools like Dialixir.

In this work our aim is to introduce a type system in order to perform static typing on expressions of both basic and structured types. Besides the mentioned basic types, our type system also manipulates types for lists, tuples, maps and functions. We also include the type \texttt{term}, the type of all terms, and \texttt{none} (the empty type). 

Our type system is based on subtyping; we write ($<:$) to denote the subtyping relation. For instance $\texttt{integer} <: \texttt{float}$. 
The type \texttt{term} is the top type in the subtyping relation.

Our type system also includes a type called \texttt{any} that is used to denote the \emph{unknown type}. This type, usual in gradual type systems, is used to represent dynamically typed fragments of code. Having two different types for representing the top and the unknown type is standard in gradual type systems \cite{ST06}. The choice of the names \texttt{term} and \texttt{any} for them has already been made in Erlang's Gradualizer \cite{gradualizer}.

Modulo subtyping, operators' typing is standard.
For example, the arithmetic operators restrict their operands to be numeric.
In our type system, this restriction is fulfilled by requiring the operands to be of a subtype of $\texttt{float}$. Then, the following are all correct expressions: 
\begin{lstlisting}
3.4 + 5.6     # float
4 + 5         # integer
4.0 + 5       # float
\end{lstlisting}
However, the following generates a type error:
\begin{lstlisting}
3 + "hi"      # wrong type
\end{lstlisting}

Following Elixir's philosophy, we are more flexible in the case of comparison operators. We allow values of different types to be compared with each other. However, the return type is $\texttt{boolean}$.
Therefore:
\begin{lstlisting}
("hi" > 5.0) or false   # boolean
("hi" > 5.0) * 3        # wrong type
\end{lstlisting}

Except for function signatures, our program codes do not contain any type annotation.
We use Elixir \texttt{@spec} directive to specify the signature of a function.
Function types are of the form $(t_1, ..., t_n) \rightarrow t$, where $t_1, ..., t_n$ are the types of the parameters and $t$ is the return type.
In the following example we define a function that takes an integer and returns a float.
\begin{lstlisting}
@spec func(integer) :: float
def func(x) do x * 42.0 end
\end{lstlisting}
Function \texttt{func} can be correctly applied to an integer:
\begin{lstlisting}
func(2)          # 84.0
\end{lstlisting}
but other kinds of applications would fail:
\begin{lstlisting}
func(2.0)        # wrong type
func("2")        # wrong type
\end{lstlisting}

In our type system, the type of a list is indexed by the type of its elements.
The following examples define correctly typed homogeneous lists:
\begin{lstlisting}
xs = [9   | []]  # [integer]
ys = [2.0 | xs]  # [float]
\end{lstlisting}
%
Heterogeneous lists can be defined as well, due to subtyping:
\begin{lstlisting}
zs = [true | ys] # [term]
\end{lstlisting}
Doing so the type information of each list element is lost. Then, the following is incorrect:
\begin{lstlisting}
[z | _] = zs 
z and true       # wrong type
\end{lstlisting}

We cannot deconstruct a list with, for instance, a tuple pattern. Thus, such patterns are wrongly typed:
\begin{lstlisting}
{x, y} = xs      # wrong type
\end{lstlisting}

In the fragment of Elixir we are considering for typing, maps behave like records. Map keys belong to the syntactic category $k$, which is composed by atoms, boolean literals and integers literals. Keys are treated like record labels and are exposed in the map type. For example, the previously defined map \texttt{m} has type $\mathtt{\%\{ :\!strange \Rightarrow string, 9 \Rightarrow boolean\}}$. Therefore, the following expression is well-typed:
\begin{lstlisting}
m[:strange] <> "bye"      # string 
\end{lstlisting}
while the following ones are not:
\begin{lstlisting}
m[:strange] + 3           # wrong type
m[10]                     # wrong type
\end{lstlisting}
Due to subtyping, pattern matching with a pattern containing a subset of the keys is also possible:
\begin{lstlisting}
case m do %{: strange => x} -> x end # string
\end{lstlisting}


One of the main objectives of the design of our type system is to be backward compatible in order to allow working with legacy code. To do so we allow the existence of \emph{untyped functions}. An untyped function is a function that does not have an associated signature (\texttt{@spec} specification). In fact, \texttt{Base.Math.dec} and \texttt{Main.fact}, previously defined, are untyped functions. We do not type check the definitions of such functions, and assign them the type $(\texttt{any}, .., \texttt{any}) \rightarrow \texttt{any}$, where $\texttt{any}$ is the unknown type. 
This type means that the function accepts arguments of any type and returns a value of an unknown type. Type correctness of both the arguments and return values in function calls to such functions is then checked at runtime.

The type $\texttt{any}$ may also be used explicitly in function signatures in order to indicate that some parameters  have dynamic types.
This provides a \emph{gradual approach} to type checking, and also makes it possible to have some sort of \emph{poor man's (parametric) polymorphism}. For example, the following function calculates the length of a list:
\begin{lstlisting}
@spec length([any]) :: integer
def length([]) do  0 end
def length([head|tail]) do 1 + length(tail) end
\end{lstlisting}
Moreover, we can define the identity function, by using $\texttt{any}$ as result type.
\begin{lstlisting}
@spec id(any) :: any
def id(x) do x end
\end{lstlisting}

The question that naturally arises is what we can do with those values (of type $\texttt{any}$) returned by untyped or gradually typed functions. 
In the places where values of type \texttt{any} occur, our type system will implicitly perform a downcast to a concrete type $t$ given by the context. This is based on the principle, usual in many OO languages, regarding the trust on downcast introductions. 
This enables us to proceed with the typing process normally, but has the well-known risk that the introduction of a downcast may lead to a runtime error. 
For example, the following expression correctly typechecks after performing a downcast from $\texttt{any}$ to $\texttt{integer}$ and runs without any problem:
\begin{lstlisting}
id(8) + 10               # 18
\end{lstlisting}
However, the following expressions typecheck, but fail at runtime:
\begin{lstlisting}
"hello" <> Main.fact(9)  # runtime error
id(8) and true           # runtime error
\end{lstlisting}

Furthermore, we can define functions that typecheck, but fail everytime they are applied:
\begin{lstlisting}
@spec bad(any) :: integer
def bad(x) do if x do x else 2 end end
\end{lstlisting}
It is important to delve into the difference between an untyped function and a gradually typed one. In the former, we perform no type check of the function body, whereas in functions like the one in this last example, a complete type check is performed, modulo the "holes" produced by downcasts on occurrences of the formal parameter \texttt{x}.

\section{Type System}\label{sec:typesystem}

In this section we present the formal components of our type system, including the subtyping relation and casting rules that allow us to perform upcasts and downcasts. We will start by defining a completely static type system, and later we will modify it in order to deal with untyped functions and gradual types.

\subsection{Static Type System}

For the static type system we assume that all functions (defined in the program or imported) come equipped with a type signature specified by a \texttt{@spec}. The type \texttt{any} is not considered for the static case. It will be added to the language when we see the gradual type system. 

\subsubsection{Program Validity}

The following judgement establishes when a program is well-typed:
\begin{equation}
    \tag{Program Validity}
    \vdash^{p} prog
    \label{j:pvalidity}
\end{equation}

\noindent
Figure~\ref{fig:programvalidity} shows the typing rules for this judgement corresponding to the cases when a program is given by a sequence of modules and expressions, respectively. 
%
%
%

In rule (P\_M), by means of the Function Signatures relation (Figure~\ref{fig:functiondeclarations}) we first traverse a module sequence $m$ in order to  
collect in an environment $\Delta$ all the function signatures that occur in it. Using the generated environment $\Delta$ we then check that the module sequence $m$ is well-typed by means of the Type Checking relation (Figure~\ref{fig:typechecking}).  
Rule (P\_E) checks that an expression sequence $e$ is well-typed by means of the Type Checking relation (Figure~\ref{fig:typechecking}). 


\begin{figure}

\begin{tabular}{c c}
\begin{minipage}{.45\columnwidth}
\begin{prooftree} 
\AxiomC{$\varnothing;\epsilon \vdash^{fs} m \Rightarrow \Delta$}
\noLine
\UnaryInfC{$\quad \Delta;\epsilon\vdash^{ch} m \quad$}
\RightLabel{(P\_M)}
\UnaryInfC{$\vdash^{p} m $}
\end{prooftree}
\end{minipage}
&
\begin{minipage}{.45\columnwidth}
\begin{prooftree} 
\AxiomC{$\quad \varnothing;\epsilon\vdash^{ch} e$}
\RightLabel{(P\_E)}
\UnaryInfC{$\vdash^{p} e $}
\end{prooftree}
\end{minipage}
\end{tabular}

\hrule
\caption{Program Validity Rules}
\label{fig:programvalidity}
\end{figure}


\subsubsection{Function Signatures}

\begin{figure}

\begin{prooftree} 
\AxiomC{$\Delta;\rho \vdash^{fs} m_{1} \Rightarrow \Delta' 
\quad \Delta';\rho \vdash^{fs} m_{2} \Rightarrow \Delta''$}
\RightLabel{(FS\_MSEQ)}
\UnaryInfC{$\Delta;\rho \vdash^{fs} m_{1};m_{2} \Rightarrow \Delta''$}
\end{prooftree}

\begin{prooftree} 
\AxiomC{$\Delta;\rho.m\_name \vdash^{fs} bdy \Rightarrow \Delta'$}
\RightLabel{(FS\_MDEF)}
\UnaryInfC{$\Delta;\rho \vdash^{fs} \texttt{defmodule} \ m\_name \ \texttt{do} \ bdy \ \texttt{end} \Rightarrow \Delta'$}
\end{prooftree}

\begin{prooftree} 
\AxiomC{$\Delta;\rho \vdash^{fs} f_{1} \Rightarrow \Delta' 
\quad \Delta';\rho \vdash^{fs} f_{2} \Rightarrow \Delta''$}
\RightLabel{(FS\_FSEQ)}
\UnaryInfC{$\Delta;\rho \vdash^{fs} f_{1};f_{2} \Rightarrow \Delta''$}
\end{prooftree}

\begin{prooftree} 
\AxiomC{$(\rho.f\!\_name, n) \not\in \Delta $}
\RightLabel{(FS\_TYSPEC)}
\UnaryInfC{$\Delta;\rho \vdash^{fs} \texttt{@spec} \ f\!\_name(t_{1},...,t_{n}) \ \texttt{::} \ t \hspace{1.5cm} $}
\noLine
\UnaryInfC{\hspace{1.5cm} $ \Rightarrow \Delta[ (\rho.f\!\_name,n) \mapsto (t_{1},...,t_{n}) \rightarrow t] $}
\end{prooftree}

\begin{prooftree} 
\AxiomC{}
\RightLabel{(FS\_FDEF)}
\UnaryInfC{$\Delta;\rho \vdash^{fs} \texttt{def[p]} \ f\!\_name (p_1,...,p_n) \ \texttt{do} \ e \ \texttt{end} \Rightarrow \Delta$}
\end{prooftree}

\begin{prooftree}
\AxiomC{}
\RightLabel{(FS\_E)}
\UnaryInfC{$\Delta;\rho \vdash^{fs} e \Rightarrow \Delta$}
\end{prooftree}

\hrule
\caption{Function signatures gathering}
\label{fig:functiondeclarations}
\end{figure}

Function signatures are processed by the rules shown in  Figure~\ref{fig:functiondeclarations}. In that process we traverse the program and collect the declared function signatures in an environment. The judgement 
\begin{equation}
    \tag{Function Signatures}
    \Delta;\rho \vdash^{fs} m \Rightarrow \Delta'
    \label{j:pdeclarations}
\end{equation}
establishes that, starting with an initial environment $\Delta$ and a module prefix $\rho$ (defining the modules hierarchy), a new environment $\Delta'$ is obtained by extending $\Delta$ with the type signatures declared in the module sequence $m$.   

In the rules (FS\_MSEQ) and (FS\_FSEQ) we collect the type signatures that occur in a sequence of modules and function declarations, respectively.  

Rule (FS\_MDEF), for a module definition, traverses the body of the module with the module prefix $\rho$ extended with the name $m\_name$ of the module; this means that all function signatures declared inside the module body will be added to the environment qualified with the extended prefix ($\rho.m\_name$).

The most interesting rule is (FS\_TYSPEC), for type specifications, where a new function signature is added to the environment. Elixir identifies functions not only by their module and name, but also by their arity. This is because it is allowed to have functions with equal name but different arity. In case the function is not in the environment $\Delta$, then we extend $\Delta$ with a new entry corresponding to the function. We use the notation $\Delta[f \mapsto s]$ for environment extension.  


Neither function definitions (FS\_FDEF) nor expressions (FS\_E) modify the environment. Notice that functions defined without associated type specifications (\texttt{@spec}) are not added to $\Delta$.

\subsubsection{Type Checking}


Figure~\ref{fig:typechecking} shows the rules for the Type Checking relation: 
\begin{equation}
    \tag{Type Checking}
    \Delta;\rho\vdash^{ch} m
    \label{j:typechecking}
\end{equation}

The rules (CH\_MSEQ) and (CH\_FSEQ) apply the type checking recursively on the subsequences. In rule (CH\_MDEF) the type checking is applied to the body of the module with the module prefix extended with $m\_name$. 

Type specifications for functions (CH\_FSPEC) are always considered correct.

\begin{figure}

\begin{tabular}{c c}
\begin{minipage}{.47\columnwidth}

\begin{prooftree} 
\AxiomC{$\Delta;\rho \vdash^{ch} m_{1}$}
\noLine
\UnaryInfC{$\Delta;\rho \vdash^{ch} m_{2}$}
\RightLabel{(CH\_MSEQ)}
\UnaryInfC{$\Delta;\rho \vdash^{ch} m_{1};m_{2}$}
\end{prooftree}
\end{minipage}
&
\begin{minipage}{.44\columnwidth}
\begin{prooftree} 
\AxiomC{$\Delta;\rho\vdash^{ch} f_{1}$}
\noLine
\UnaryInfC{$\Delta;\rho\vdash^{ch} f_{2}$}
\RightLabel{(CH\_FSEQ)}
\UnaryInfC{$\Delta;\rho\vdash^{ch} f_{1};f_{2}$}
\end{prooftree}
\end{minipage}
\end{tabular}

\begin{prooftree} 
\AxiomC{$\Delta;\rho.m\_name \vdash^{ch} bdy $}
\RightLabel{(CH\_MDEF)}
\UnaryInfC{$\Delta;\rho \vdash^{ch} \texttt{defmodule} \ m\_name \ \texttt{do} \ bdy \ \texttt{end}$}
\end{prooftree}

\begin{prooftree} 
\AxiomC{ }
\RightLabel{(CH\_FSPEC)}
\UnaryInfC{$\Delta;\rho \vdash^{ch} \texttt{@spec} \ f\!\_name(t_{1},...,t_{n}) \ \texttt{::} \ t$}
\end{prooftree}

\begin{prooftree} 
\AxiomC{$\Delta(\rho.f\!\_name, n) = t_{1},...,t_{n} \rightarrow t $}
\noLine
\UnaryInfC{$\varnothing;\varnothing \vdash^{tp} p_{1} : t_{1} \Rightarrow \Gamma^1  \quad ... \quad   \varnothing;\Gamma^{n-1} \vdash^{tp} p_{n} : t_{n} \Rightarrow \Gamma^n $}
\noLine
\UnaryInfC{$\Delta; \Gamma^n; \rho \vdash^t e : t \Rightarrow \Gamma'$}
\RightLabel{(CH\_FDEFT)}
\UnaryInfC{$\Delta; \rho \vdash^{ch} \texttt{def} \ f\!\_name (p_{1},...,p_{n}) \ \texttt{do} \ e \ \texttt{end}$}
\end{prooftree}

\begin{prooftree} 
\AxiomC{$\quad \Delta;\varnothing;\rho\vdash^{t} e : t \Rightarrow \Gamma$}
\RightLabel{(CH\_E)}
\UnaryInfC{$\Delta;\rho\vdash^{ch} e $}
\end{prooftree}

\hrule
\caption{Type Checking Rules}
\label{fig:typechecking}
\end{figure}

%
Rule (CH\_FDEFT) considers function definitions. In those cases we check that the patterns $p_i$ have type $t_i$ and the function body $e$ has type $t$, where $(t_1,\ldots,t_n) \rightarrow t$ is the declared function type contained in the environment. During the type checking of the patterns $p_i$ we incrementally collect their variables in an environment that finishes with $\Gamma^n$, which is then used to type check the function body $e$. Elixir allows multiple occurrences of a variable in patterns. In that case all occurrences of the variable must bind to the same value. The type checking of patterns (Figure~\ref{fig:patterntyping}) requires that all such occurrences must have exactly the same type. The following is an example of an incorrect definition:
\begin{lstlisting}
@spec func(integer, string) :: integer
def func(x, x) do x end 
\end{lstlisting}
while the following is an example of a well-typed one:
\begin{lstlisting}
@spec func(integer, integer) :: integer
def func(x, x) do x end    
\end{lstlisting}


Rule (CH\_E) establishes that an expression $e$ type checks when we can deduce it is well-typed for some type $t$. 



\subsubsection{Pattern Rules}

\begin{figure}

\begin{tabular}{c c}
\begin{minipage}{0.45\columnwidth}
\begin{prooftree} 
\AxiomC{$ $}
\RightLabel{(TP\_WILD)}
\UnaryInfC{$\Sigma; \Gamma \vdash^{tp} \_ : t \Rightarrow \Gamma  $}
\end{prooftree}
\end{minipage}
&
\begin{minipage}{0.47\columnwidth}
\begin{prooftree} 
\AxiomC{$t = ty(l)$}
\RightLabel{(TP\_LIT)}
\UnaryInfC{$\Sigma;\Gamma; \rho \vdash^{tp} l : t \Rightarrow \Gamma$}
\end{prooftree}
\end{minipage}
\end{tabular}

\begin{tabular}{c c}
\begin{minipage}{0.46\columnwidth}
\begin{prooftree} 
\AxiomC{$ \Gamma[x] = t$}
\RightLabel{(TP\_VARE)}
\UnaryInfC{$\Sigma;\Gamma \vdash^{tp} x: t \Rightarrow \Gamma$}
\end{prooftree}
\end{minipage}
&
\begin{minipage}{0.51\columnwidth}
\begin{prooftree} 
\AxiomC{$ x \not\in \Gamma $}
\RightLabel{(TP\_VARN)}
\UnaryInfC{$\Sigma;\Gamma \vdash^{tp} x: t \Rightarrow \Gamma[ x \mapsto t]$}
\end{prooftree}
\end{minipage}
\end{tabular}

\begin{prooftree} 
\AxiomC{$ \Sigma[x] = t$}
\RightLabel{(TP\_PIN)}
\UnaryInfC{$\Sigma;\Gamma \vdash^{tp} \texttt{\^{}} x: t \Rightarrow \Gamma$}
\end{prooftree}



\begin{prooftree} 
\AxiomC{$\Sigma;\Gamma \vdash^{tp} p_{1}:t_{1} \Rightarrow \Gamma^1 
\quad ...
\quad \Sigma;\Gamma^{n-1} \vdash^{tp} p_{n}:t_{n} \Rightarrow \Gamma^n $}
\RightLabel{(TP\_TUP)}
\UnaryInfC{$\Sigma;\Gamma \vdash^{tp} \{p_{1}, ..., p_{n} \} : \{t_{1}, ..., t_{n}\} \Rightarrow \Gamma^n$}
\end{prooftree}

\begin{prooftree} 
\AxiomC{}
\RightLabel{(TP\_ELIST)}
\UnaryInfC{$\Sigma;\Gamma \vdash^{tp} [\ ]: [t] \Rightarrow \Gamma $}
\end{prooftree}

\begin{prooftree} 
\AxiomC{$\Sigma;\Gamma \vdash^{tp} p_{1}:t \Rightarrow \Gamma' 
\quad \Sigma;\Gamma' \vdash^{tp} p2: [t] \Rightarrow \Gamma'' $}
\RightLabel{(TP\_LIST)}
\UnaryInfC{$\Sigma;\Gamma \vdash^{tp} [p_{1} |\ p2] : [t] \Rightarrow \Gamma'' $}
\end{prooftree}

\begin{prooftree} 
\AxiomC{}
\RightLabel{(TP\_EMAP)}
\UnaryInfC{$\Sigma;\Gamma \vdash^{tp} \%\{\ \} : \%\{\ \} \Rightarrow \Gamma $}
\end{prooftree}

\begin{prooftree} 
\AxiomC{$\Sigma;\Gamma \vdash^{tp} p_{1}:t_{1} \Rightarrow \Gamma^1 
\quad ... 
\quad \Sigma;\Gamma^{n-1} \vdash^{tp} p_{n}:t_{n} \Rightarrow \Gamma^n $}
\RightLabel{(TP\_MAP)}
\UnaryInfC{$\Sigma;\Gamma \vdash^{tp} \%\{k_{1} \Rightarrow p_{1},..., k_{n} \Rightarrow p_{n}\} $}
\noLine
\UnaryInfC{$\hspace{30mm} 
: \%\{\ k_1 \Rightarrow t_{1}, ..., k_n \Rightarrow t_{n} \} \Rightarrow \Gamma^n $}
\end{prooftree}

\hrule
\caption{Pattern Typing Rules}
\label{fig:patterntyping}
\end{figure}
Patterns are type checked using the following judgement:
\begin{equation}
    \tag{Pattern Typing}
    \Sigma;\Gamma \vdash^{tp} p : t \Rightarrow \Gamma'
    \label{j:patterntyping}
\end{equation}
It expresses that a pattern $p$ has type $t$ in the variable environments $\Sigma$ (global) and $\Gamma$ (local). The environment $\Sigma$ contains the variable definitions in the context of the pattern while $\Gamma$ contains the variables that are locally defined in the pattern. A new environment $\Gamma'$ is produced by extending $\Gamma$ with the new variables defined in the pattern. The typing rules for patterns are shown in Figure~\ref{fig:patterntyping}.
The need to maintain two variable environments lies in the fact that we have to deal with pinned variables (that are bound in the global environment) and non-linear patterns (where multiple occurrences of variables within a pattern have to be checked to have the same type).

The wild pattern \texttt{\_} matches with any value. Therefore, the rule (TP\_WILD) accepts any type $t$ and does not extend the environment.

Literal patterns (TP\_LIT) also do not extend the environment, and have the corresponding type given by: 
\begin{quote}
$\begin{array}{l@{\;}l l l@{\;}l } 
ty(i) & = \texttt{integer} & \; & ty(s) &  = \texttt{string} \\
ty(r) & = \texttt{float} & \; & ty(b) & = \texttt{boolean} \\
ty(a) & = a & & \\
\end{array}$
\end{quote}
Notice that atoms are singleton types.

The rule (TP\_VARN), for newly introduced variables, extends the environment with the variable $x$ associated to the type $t$ of the pattern. 
On the other hand, when the variable is already present in the environment $\Gamma$, which is possible because patterns are non-linear, by rule (TP\_VARE) its type must coincide with the type in $\Gamma$.
The rule (TP\_PIN), for pinned variables, is similar to (TP\_VARE), but in this case the type of the variable is taken from the global environment $\Sigma$.

Rules (TP\_TUP), (TP\_ELIST, TP\_LIST) and (TP\_EMAP, TP\_MAP), for tuple, list and map patterns, respectively, are quite straightforward. In all of them we collect the variables that appear in the patterns; the same requirement regarding the multiple occurrence of variables in patterns holds here (i.e. multiple occurrences of a variable must have the same type). 

\subsubsection{Subtyping}

Our type system is based on subtyping~\cite{pierce}. As usual, the subtyping relation, noted as $<:$, is used to represent set inclusion, where types are interpreted as collections of values.
An element of one type is then considered an element of its supertypes.

\begin{figure}

\begin{tabular}{c c}
\begin{minipage}{.35\columnwidth}
\begin{prooftree} 
\AxiomC{ }
\RightLabel{(ST\_REFL)}
\UnaryInfC{$t <: t$}
\end{prooftree}
\end{minipage}
&
\begin{minipage}{.50\columnwidth}
\begin{prooftree} 
\AxiomC{$ t_{1} <: t_{2} \quad t_{2} <: t_{3}$}
\RightLabel{(ST\_TRANS)}
\UnaryInfC{$t_{1} <: t_{3}$}
\end{prooftree}
\end{minipage}
\end{tabular}

\begin{prooftree} 
\AxiomC{$ u_{1} <: t_{1} \ \ \cdots \ \ u_{n} <: t_{n} \quad\quad t <: u$}
\RightLabel{(ST\_FUN)}
\UnaryInfC{$(t_{1},...,t_{n}) \rightarrow t <: (u_{1},...,u_{n}) \rightarrow u$}
\end{prooftree}

\vspace{6pt}

\begin{tabular}{c c}
\begin{minipage}{.42\columnwidth}
\begin{prooftree} 
\AxiomC{ }
\RightLabel{(ST\_TERM)}
\UnaryInfC{$t <: \texttt{term}$}
\end{prooftree}
\end{minipage}
&
\begin{minipage}{.42\columnwidth}
\begin{prooftree} 
\AxiomC{ }
\RightLabel{(ST\_NONE)}
\UnaryInfC{$\texttt{none} <: t$}
\end{prooftree}
\end{minipage}\\
\end{tabular}

\begin{tabular}{c c}
\begin{minipage}{.46\columnwidth}
\begin{prooftree} 
\AxiomC{ }
\RightLabel{(ST\_NUM)}
\UnaryInfC{$\texttt{integer} <: \texttt{float}$}
\end{prooftree}
\end{minipage}
&
\begin{minipage}{.46\columnwidth}
\begin{prooftree} 
\AxiomC{}
\RightLabel{(ST\_ATOM)}
\UnaryInfC{$a <: \texttt{atom}$}
\end{prooftree}
\end{minipage}
\end{tabular}

\begin{tabular}{c c}
\begin{minipage}{.30\columnwidth}
\begin{prooftree} 
\AxiomC{$t <: u$}
\RightLabel{(ST\_LIST)}
\UnaryInfC{$[t] <: [u]$}
\end{prooftree}
\end{minipage}
&
\begin{minipage}{.60\columnwidth}
\begin{prooftree} 
\AxiomC{$t_{i} <: u_{i} \quad \forall i \in 1,..,n$}
\RightLabel{(ST\_TUPLE)}
\UnaryInfC{$\{t_{1},..,t_{n}\} <: \{u_{1},..., u_{n}\}$}
\end{prooftree}
\end{minipage}
\end{tabular}

\begin{prooftree} 
\AxiomC{$t_{i} <: u_{i} \quad \forall i \in 1,..,n$}
\RightLabel{(ST\_MAP)}
\UnaryInfC{$\%\{k_1 \Rightarrow t_{1},.., k_{n+m} \Rightarrow t_{n+m}\} <: \%\{k_1 \Rightarrow u_{1},.., k_{n} \Rightarrow u_{n}\}$}
\end{prooftree}

\hrule
\caption{Subtyping Relation}
\label{fig:subtyping}
\end{figure}

Figure~\ref{fig:subtyping} shows the properties of the subtyping relation. The subtyping relation is reflexive (ST\_REFL) and transitive (ST\_TRANS). 
The type \texttt{none} is subtype of all types (ST\_NONE), whereas the type \texttt{term} is supertype of all types (ST\_TERM). Concerning the numeric types, (ST\_NUM) states that \texttt{integer} is subtype of \texttt{float}.
In rule (ST\_ATOM) we say that all the atoms are subtypes of \texttt{atom}.
For lists, rule (ST\_LIST) states that the list type constructor preserves the subtyping relation of the element types. The subtyping rule for tuples (ST\_TUPLE) works componentwise. 
Rule (ST\_MAP) is similar to the usual subtyping rule for record types. A map with less key-value pairs (i.e. with fewer $u_i$ types) is supertype of a map with more entries, provided that the types of the entries shared by the two map types are in covariant relation.
Since order does not matter, we consider that a map is subtype of any of its permutations.
Figure~\ref{fig:expression:sub} shows the subsumption rule (ST\_SUB).

\begin{figure}

\begin{prooftree} 
\AxiomC{$\Delta; \Gamma; \rho \vdash^{t} e : t \Rightarrow \Gamma' 
\quad t <: u $} 
\RightLabel{(T\_SUB)}
\UnaryInfC{$\Delta; \Gamma; \rho \vdash^{t} e : u \Rightarrow \Gamma' $}
\end{prooftree}

\hrule
\caption{Expression subsumption}
\label{fig:expression:sub}
\end{figure}

\subsubsection{Expression Typing}


Expressions are type checked using the following judgement:
\begin{equation}
    \tag{Expression Typing}
    \Delta;\Gamma;\rho\vdash^{t} e : t \Rightarrow \Gamma'
    \label{j:expressiontyping}
\end{equation}
It states that an expression $e$ has type $t$ in a function environment $\Delta$, a variable environment $\Gamma$, and a module prefix $\rho$. A new environment $\Gamma'$ is produced by extending $\Gamma$ with the new variables defined in $e$.




\begin{figure}
\begin{minipage}{.49\columnwidth}
\begin{prooftree} 
\AxiomC{$t = ty(l)$}
\RightLabel{(T\_LIT)}
\UnaryInfC{$\Delta; \Gamma; \rho \vdash^{t} l : t \Rightarrow \Gamma $}
\end{prooftree}
\end{minipage}
\begin{minipage}{.49\columnwidth}
\begin{prooftree} 
\AxiomC{$\Gamma(x) = t $}
\RightLabel{(T\_VAR)}
\UnaryInfC{$\Delta; \Gamma; \rho \vdash^{t} x: t \Rightarrow \Gamma$}
\end{prooftree}
\end{minipage}

\hrule
\caption{Typing of Literals and Variables}
\label{fig:expression:literals}
\end{figure}

Figure~\ref{fig:expression:literals} shows the rules for typing literals and variables. Literals (T\_LIT) have the type determined by $ty$, defined before. Variables (T\_VAR) have the type with which they were bound in the environment $\Gamma$.

\begin{figure}

\begin{prooftree} 
\AxiomC{$\Delta;\Gamma;\rho \vdash^{t} e : t \Rightarrow \Gamma' 
\quad t <: \texttt{float} $}
\RightLabel{(T\_NEG)}
\UnaryInfC{$\Delta;\Gamma;\rho \vdash^{t} \texttt{-} e : t \Rightarrow \Gamma' $}
\end{prooftree}

\begin{prooftree} 
\AxiomC{$\diamond \in \{+,-,*\} \quad t <: \texttt{float} $}
\noLine
\UnaryInfC{$\Delta;\Gamma;\rho \vdash^{t} e_1 : t \Rightarrow \Gamma^1 
\quad\Delta;\Gamma;\rho \vdash^{t} e_2 : t \Rightarrow \Gamma^2 $}
\RightLabel{(T\_ARITH)}
\UnaryInfC{$\Delta;\Gamma;\rho \vdash^{t} e_1 \diamond e_2 : t \Rightarrow \Gamma^1 \dagger \Gamma^2 $}
\end{prooftree}

\begin{prooftree} 
\AxiomC{$\Delta;\Gamma;\rho \vdash^{t} e_1 : \texttt{float} \Rightarrow \Gamma^1 
\quad\Delta;\Gamma;\rho \vdash^{t} e_2 : \texttt{float} \Rightarrow \Gamma^2 $}
\RightLabel{(T\_DIV)}
\UnaryInfC{$\Delta;\Gamma;\rho \vdash^{t} e_1 \texttt{/} e_2 : \texttt{float} \Rightarrow \Gamma^1 \dagger \Gamma^2 $}
\end{prooftree}
\hrule
\caption{Typing of Arithmetic Operators}
\label{fig:expression:arith}
\end{figure}

\begin{figure}
\begin{prooftree} 
\AxiomC{$\Delta;\Gamma;\rho \vdash^{t} e : \texttt{boolean} \Rightarrow \Gamma' $}
\RightLabel{(T\_NOT)}
\UnaryInfC{$\Delta;\Gamma;\rho \vdash^{t} \texttt{not} \ e : \texttt{boolean} \Rightarrow \Gamma' $}
\end{prooftree}

\begin{prooftree}
\AxiomC{$\bullet \in \{and,or\}$}
\noLine
\UnaryInfC{$\Delta;\Gamma;\rho \vdash^{t} e_1 : \texttt{boolean} \Rightarrow \Gamma^1 $}
\noLine
\UnaryInfC{$\Delta;\Gamma;\rho \vdash^{t} e_2 : \texttt{boolean} \Rightarrow \Gamma^2 $}
\RightLabel{(T\_BOP)}
\UnaryInfC{$\Delta;\Gamma;\rho \vdash^{t} e_1 \bullet e_2 : \texttt{boolean} \Rightarrow \Gamma^1 \dagger \Gamma^2 $}
\end{prooftree}

\begin{prooftree} 
\AxiomC{$\star \in \{\texttt{<,>,<=,>=,==,!=,===,!==}\}$}
\noLine
\UnaryInfC{$
\Delta;\Gamma;\rho \vdash^{t} e_1 : t_1 \Rightarrow \Gamma^1 
\quad \Delta;\Gamma;\rho \vdash^{t} e_2 : t_2 \Rightarrow \Gamma^2 $}
\RightLabel{(T\_CMP)}
\UnaryInfC{$\Delta;\Gamma;\rho \vdash^{t} e_1 \star e_2 : \texttt{boolean} \Rightarrow \Gamma^1 \dagger \Gamma^2 $}\\
\end{prooftree}
\hrule
\caption{Typing of Boolean Operators}
\label{fig:expression:bool}
\end{figure}

\begin{figure}
\begin{prooftree} 
\AxiomC{$\square \in \{++,--\}$}
\noLine
\UnaryInfC{$\Delta;\Gamma;\rho \vdash^{t} e_1 : [t] \Rightarrow \Gamma^1 $}
\noLine
\UnaryInfC{$\Delta;\Gamma;\rho \vdash^{t} e_2 : [t] \Rightarrow \Gamma^2 $}
\RightLabel{(T\_LOP)}
\UnaryInfC{$\Delta;\Gamma;\rho \vdash^{t} e_1 \,\square\, e_2 : [t] \Rightarrow \Gamma^1 \dagger \Gamma^2 $}
\end{prooftree}

\begin{prooftree} 
\AxiomC{$\Delta;\Gamma;\rho \vdash^{t} e_1 : \texttt{string} \Rightarrow \Gamma^1 $}
\noLine
\UnaryInfC{$\Delta;\Gamma;\rho \vdash^{t} e_2 : \texttt{string} \Rightarrow \Gamma^2 $}
\RightLabel{(T\_CONCAT)}
\UnaryInfC{$\Delta;\Gamma;\rho \vdash^{t} e_1 \, \texttt{\textless}\texttt{\textgreater} \, e_2 : \texttt{string} \Rightarrow \Gamma^1 \dagger \Gamma^2 $}
\end{prooftree}

\hrule
\caption{Typing of List and String Operators}
\label{fig:expression:list}
\end{figure}

Figures \ref{fig:expression:arith}, \ref{fig:expression:bool} and \ref{fig:expression:list} show the rules for operators. In the case of negation (T\_NEG) and arithmetic operations (T\_ARITH), we impose the type $t$ of the expression to be numeric by restricting it to be a subtype of $\texttt{float}$. For instance, we can type \texttt{-9}:

\begin{prooftree} 
\AxiomC{$\texttt{integer} = ty(9)$}
\RightLabel{\scriptsize{(T\_LIT)}}
\UnaryInfC{$\varnothing;\varnothing;\epsilon \vdash^{t} \texttt{9} : \texttt{integer} \Rightarrow \varnothing$}
\AxiomC{$\texttt{integer} <: \texttt{float}$}
\RightLabel{\scriptsize{(T\_NEG)}}
\BinaryInfC{$\varnothing;\varnothing;\epsilon \vdash^{t} \texttt{-9} : \texttt{integer} \Rightarrow \varnothing $}
\end{prooftree}

The operator $\Gamma^1 \dagger \Gamma^2$, used in (T\_ARITH) and many other rules, performs the union of both environments, choosing the components of $\Gamma^2$ in case of overlapping. 
Thus, if both operands introduce variables with the same name, the ones introduced by the right operand will remain in scope. This is the behaviour in all the cases of binary operators, as can be seen in the corresponding rules.

Division accepts operands of subtypes of $\texttt{float}$, but always returns a $\texttt{float}$. This is the only difference between rule (T\_DIV) and the rule (T\_ARITH) for the other numeric operators. 
In the following derivation, we can see the use of subsumption when typing a division with an integer operand:

\begin{prooftree} 
\def\defaultHypSeparation{\hskip .1in}
\def\ScoreOverhang{1pt}
\AxiomC{$...$}
\AxiomC{$\texttt{integer} = ty(2)$}
\LeftLabel{\scriptsize{(T\_LIT)}}
\UnaryInfC{$\varnothing;\varnothing;\epsilon \vdash^{t} \texttt{2} : \texttt{integer} \Rightarrow \varnothing $}
\AxiomC{$\texttt{integer} <: \texttt{float}$}
\RightLabel{\scriptsize{(T\_SUB)}}
\BinaryInfC{$\varnothing;\varnothing;\epsilon \vdash^{t} \texttt{2} : \texttt{float} \Rightarrow \varnothing $}
\RightLabel{\scriptsize{(T\_DIV)}}
\BinaryInfC{$\varnothing;\varnothing;\epsilon \vdash^{t} \texttt{9.0 / 2} : \texttt{float} \Rightarrow \varnothing $}
\end{prooftree}

\begin{figure}

\begin{prooftree} 
\AxiomC{$\Delta; \Gamma ; \rho \vdash^t e:t \Rightarrow \Gamma^1 
\quad \Gamma;\varnothing \vdash^{tp} p:t \Rightarrow \Gamma^2$}
\RightLabel{(T\_MATCH)}
\UnaryInfC{$\Delta; \Gamma; \rho \vdash^{tp} p = e : t  \Rightarrow \Gamma^1 \dagger \Gamma^2$}
\end{prooftree}

\begin{prooftree} 
\AxiomC{$\Gamma;\varnothing \vdash^{tp} p_{1} : t_{1} \Rightarrow \Gamma^1  \quad ... \quad   \Gamma;\Gamma^{n-1} \vdash^{tp} p_{n} : t_{n} \Rightarrow \Gamma^n $}
\noLine
\UnaryInfC{$\Delta; \Gamma \dagger \Gamma^n; \rho \vdash^t e : t \Rightarrow \Gamma'$}
\RightLabel{(T\_ANON)}
\UnaryInfC{$\Delta; \Gamma; \rho \vdash^{t} \texttt{fn} \ (p_{1},...,p_{n}) \ \texttt{->} \ e \ \texttt{end} : (t_1,...,t_n) \rightarrow \  t \Rightarrow \Gamma'$}
\end{prooftree}

\begin{prooftree} 
\AxiomC{$\Delta;\Gamma;\rho \vdash^{t} e_{1} : t_{1} \Rightarrow \Gamma' 
\quad \Delta;\Gamma';\rho \vdash^{t} e_{2} : t_{2} \Rightarrow \Gamma'' $}
\RightLabel{(T\_ES)}
\UnaryInfC{$\Delta;\Gamma;\rho \vdash^{t} e_{1};e_{2} : t_{2} \Rightarrow \Gamma'' $}
\end{prooftree}

\hrule
\caption{Typing of Matchings, Anonymous Functions and Sequences}
\label{fig:expression:bindseq}
\end{figure}

Figure~\ref{fig:expression:bindseq} introduces the rules for matchings, anonymous functions and sequences. In rule (T\_MATCH), the type $t$ of the expression has to be a subtype of the type $u$ of the pattern. This lets us to type, for example, the following expression, where we bind the value $1$ to the variable $x$:
\begin{lstlisting}
%{:yes => x} = %{:yes => 1, :no => 2} 
\end{lstlisting}

Both the pattern and the expression can introduce new (or rebind existing) variables. The environment $\Gamma^1$ results from extending the current environment $\Gamma$ with the variables introduced by $e$, while $\Gamma^2$ contains the variables introduced by the pattern $p$. The variables introduced by the pattern have greater priority than those introduced in the expression.

Rule (T\_ANON), for checking the definition of anonymous functions, is similar to (CH\_FDEFT) of Figure~\ref{fig:typechecking}, although in this case there is no signature in $\Delta$ to check with it. The global environment $\Sigma$ is set with the environment $\Gamma$ because it is possible to use pinned variables within patterns of anonymous functions. For example,
\begin{lstlisting}
x = 3
f = fn(^x,y) -> x + y end
f.(2,1)                      # matching error
\end{lstlisting}

In the case of a sequence (T\_ES), we type the sub-expressions weaving the extensions of the environment from left to right. Notice that the type of a sequence is the type of its rightmost sub-expression.

\begin{figure}

\begin{prooftree} 
\AxiomC{$\Delta;\Gamma;\rho \vdash^{t} e : \texttt{boolean} \Rightarrow \Gamma' $}
\noLine
\UnaryInfC{$\Delta;\Gamma';\rho \vdash^{t} e_{1} : t \Rightarrow \Gamma^1 
\quad \Delta;\Gamma';\rho \vdash^{t} e_{2} : t \Rightarrow \Gamma^2 $}
\RightLabel{(T\_IFELSE)}
\UnaryInfC{$\Delta;\Gamma;\rho \vdash^{t} \texttt{if} \ e \ \texttt{do} \ e_{1} \ \texttt{else} \ e_{2} \ \texttt{end} : t \Rightarrow \Gamma' $}
\end{prooftree}


\begin{prooftree} 
\AxiomC{$\Delta;\Gamma;\rho \vdash^{t} e : t \Rightarrow \Gamma' $}
\noLine
\UnaryInfC{$\Gamma';\varnothing \vdash^{tp} p_{i} : t_i \Rightarrow \Gamma^i \quad t_i <: t \quad \forall i \in 1,..,n$}
\noLine
\UnaryInfC{$\Delta;\Gamma' \dagger \Gamma^i;\rho \vdash^{t} e_{i} : u \Rightarrow \Gamma'^{i} \quad \forall i \in 1,..,n$}
\RightLabel{(T\_CASE)}
\UnaryInfC{$\Delta;\Gamma;\rho \vdash^{t} \texttt{case} \ e \ \texttt{do} \ p_{1} \rightarrow e_{1} ... p_{n} \rightarrow e_{n} \ \texttt{end} : u \Rightarrow \Gamma' $}
\end{prooftree}

\begin{prooftree} 
\AxiomC{$\Delta;\Gamma;\rho \vdash^{t} e_{i} : \texttt{boolean} \Rightarrow \Gamma^i \quad \forall i \in 1,..,n$}
\noLine
\UnaryInfC{$\Delta;\Gamma^i;\rho \vdash^{t} e'_{i} : t \Rightarrow \Gamma'^{i}  \quad \forall i \in 1,..,n$}
\RightLabel{(T\_CND)}
\UnaryInfC{$\Delta;\Gamma;\rho \vdash^{t} \texttt{cond} \ \texttt{do} \ e_{1} \rightarrow e'_{1},...,e_{n} \rightarrow e'_{n} \ \texttt{end} : t \Rightarrow \Gamma $}
\end{prooftree}

\hrule
\caption{Typing of Control Structures}
\label{fig:expression:control}
\end{figure}

Figure~\ref{fig:expression:control} shows the rules for the control structures. Note that in all cases the bindings of the different branches do not escape outside the expression.
This is the actual behaviour of Elixir. For instance, the following sequence evaluates to the tuple \texttt{\{3,"bye"\}}:
\begin{lstlisting}
y = if (x=3) > 2 do x="bye" end
{x,y}                                   # {3,"bye"}
\end{lstlisting}

In rule (T\_CASE), we type the i-th branch of the \texttt{case} expression using the union of the environment $\Gamma'$, that results from typing the selector $e$, with the environment $\Gamma^i$ that contains the variables bound by the pattern $p_i$.
The following is a fragment of the derivation for a \texttt{case} with two different atoms in the patterns of the branches. 

\begin{prooftree} 
\def\defaultHypSeparation{\hskip .1in}
\def\ScoreOverhang{1pt}
\AxiomC{$\vdash^{t}\texttt{:yes} : \texttt{:yes}$}
\RightLabel{\scriptsize{(T\_SUB)}}
\UnaryInfC{$\vdash^{t}\texttt{:yes} : \texttt{atom} $}
\AxiomC{$\vdash^{t}\texttt{:yes} : \texttt{:yes} \quad \texttt{:yes} <: \texttt{atom}$}
\noLine
\UnaryInfC{$\vdash^{t}\texttt{:no} : \texttt{:no} \quad \texttt{:no} <: \texttt{atom}$} 
\AxiomC{$...$}
\RightLabel{\scriptsize{(T\_CASE)}}
\TrinaryInfC{$\vdash^{t} \texttt{case :yes do :yes} \rightarrow 1 ;  \texttt{:no} \rightarrow 2 :  \texttt{integer}$}
\end{prooftree}
%
Notice the use of subsumption to upcast the type of \texttt{:yes} to \texttt{:atom} in order to become a supertype of the types of the patterns. 
In fact, due to subsumption, we can upcast the type of the selector $e$ of a \texttt{case} even till \texttt{term}, allowing the patterns of the branches to have any type. We have decided to maintain this flexibility in our type system, which in turn coincides with the spirit of Elixir programming. This does not break the type safety of the language. In the worst case we will find \texttt{case} expressions with some patterns that never match the selector.


\begin{figure}

\begin{prooftree} 
\AxiomC{}
\RightLabel{(T\_ELIST)}
\UnaryInfC{$\Delta;\Gamma; \rho \vdash^{t} [\ ]: [t] \Rightarrow \Gamma $}
\end{prooftree}

\begin{prooftree} 
\AxiomC{$
\Delta; \Gamma; \rho \vdash^{t} e_{1}:t \Rightarrow \Gamma^1
\quad \Delta; \Gamma; \rho \vdash^{t} e2: [t] \Rightarrow \Gamma^2$}
\RightLabel{(T\_LIST)}
\UnaryInfC{$\Delta; \Gamma; \rho \vdash^{t} [e_{1} |\ e2] : [t] \Rightarrow \Gamma^1 \dagger \Gamma^2 $} 
\end{prooftree}

\begin{prooftree} 
\AxiomC{$\Delta; \Gamma; \rho \vdash^{t} e_{i}:t_{i} \Rightarrow \Gamma^i \quad \forall i \in 1,..n $}
\RightLabel{(T\_TUP)}
\UnaryInfC{$\Delta; \Gamma; \rho \vdash^{t} \{e_{1}, ..., e_{n} \} : \{t_{1}, ..., t_{n}\} \Rightarrow \Gamma^1 \dagger ... \dagger \Gamma^n$}
\end{prooftree}

\begin{prooftree} 
\AxiomC{}
\RightLabel{(T\_EMAP)}
\UnaryInfC{$\Delta;\Gamma;\rho \vdash^{t} \%\{\ \} : \%\{\ \} \Rightarrow \Gamma $}
\end{prooftree}

\begin{prooftree} 
\AxiomC{$\Delta; \Gamma; \rho \vdash^{t} e_{i}: t_i \Rightarrow \Gamma^{i} 
\quad \forall i \in 1,..,n $}
\RightLabel{(T\_MAP)}
\UnaryInfC{$\Delta; \Gamma; \rho \vdash^{t} \%\{k_{1} \Rightarrow e_{1},..., k_{n} \Rightarrow e_{n}\} : $}
\noLine
\UnaryInfC{$\hspace{10mm}\%\{k_1 \Rightarrow t_{1}, ..., k_n \Rightarrow t_{n}\} \Rightarrow \Gamma^{1} \dagger ... \dagger \Gamma^{n}$}
\end{prooftree}

\begin{prooftree} 
\AxiomC{$\Delta; \Gamma; \rho \vdash^{t} e : \%\{k \Rightarrow t\} \Rightarrow \Gamma' $}
\RightLabel{(T\_MAPAPP)}
\UnaryInfC{$\Delta; \Gamma; \rho \vdash^{t} e [k] : t \Rightarrow \Gamma' $} 
\end{prooftree}

\hrule
\caption{Typing of Data Structures}
\label{fig:expression:data}
\end{figure}

In Figure~\ref{fig:expression:data} we introduce the typing rules for list, tuple and map constructors, and for map application. 

Lists are homogeneous structures with type $[t]$; thus, as expressed in rule (T\_LIST), the head and tail expressions must have type $t$ and $[t]$, respectively. When typing the empty list (T\_ELIST) any type $t$ derived from the context is correct.

In (T\_MAP), we see that the type of a non-empty map exposes its keys and the type of the values associated with each of them;
keys are restricted to literals. 
In a map application (T\_MAPAPP) we only allow the application to a literal (and not to an arbitrary expression) 
in order to be able to statically determine the field that is selected and therefore the type of the result. 
For example,
\begin{lstlisting}
m = %{:good => 9, :bad => "ups"} 
m[:good] + 1                # 10
m[:bad] + 1                 # wrong
\end{lstlisting}
%


\begin{figure}

\begin{prooftree} 
\AxiomC{$\Delta(\rho_{2}.f\!\_name, n) = (t_{1},...,t_{n}) \rightarrow t $}
\noLine
\UnaryInfC{$\Delta; \Gamma; \rho_{1} \vdash^t e_{i}: t_{i} \Rightarrow \Gamma^i \quad  \forall i \in 1,..,n $}
\RightLabel{(T\_NAPPT)}
\UnaryInfC{$\Delta; \Gamma; \rho_{1} \vdash^t \rho_{2}.f\!\_name \ (e_{1},..., e_{n}) : t \Rightarrow \Gamma^1 \dagger ... \dagger \Gamma^n$}
\end{prooftree}

\begin{prooftree} 
\AxiomC{$\Delta(\rho.f\!\_name, n) = (t_{1},...,t_{n}) \rightarrow t $}
\noLine
\UnaryInfC{$\Delta; \Gamma; \rho \vdash^t e_{i}: t_{i} \Rightarrow \Gamma^i \quad \forall i \in 1,..,n $}
\RightLabel{(T\_LAPPT)}
\UnaryInfC{$\Delta; \Gamma; \rho \vdash^t f\!\_name \ (e_{1},..., e_{n}) : t \Rightarrow \Gamma^1 \dagger ... \dagger \Gamma^n$}
\end{prooftree}

\begin{prooftree} 
\AxiomC{$\Gamma(x) = (t_{1},...,t_{n}) \rightarrow t $}
\noLine
\UnaryInfC{$\Delta; \Gamma; \rho \vdash^t e_{i}: t_{i} \Rightarrow \Gamma^i \quad \forall i \in 1,..,n $}
\RightLabel{(T\_VAPPT)}
\UnaryInfC{$\Delta; \Gamma; \rho \vdash^t x \texttt{.} (e_{1},..., e_{n}) : t \Rightarrow \Gamma^1 \dagger ... \dagger \Gamma^n$}
\end{prooftree}

\hrule
\caption{Typing rules for function application}
\label{fig:expression:function}
\end{figure}


Figure~\ref{fig:expression:function} shows the rules for function application. Functions can be either local or non-local to the module where they are being called. In the function environment $\Delta$ we store the names of the functions prefixed by the module hierarchy they belong to.
Since the name of non-local functions occur prefixed by their own module hierarchy $\rho_2$, in rule (T\_NAPPT) we just use the prefixed name to determine if they belong to $\Delta$. 
In the case of local functions, in rule (T\_LAPPT) we have to add $\rho$ (the current point in the module hierarchy) as prefix to the name of the function in order to find it in the environment $\Delta$.
Rules (T\_NAPPT) and (T\_LAPPT) apply when the function belongs to the environment $\Delta$ and has type $(t_1,..., t_n) \rightarrow t$. Notice that, because of subsumption, the type of each argument $e_i$ can be a subtype of $t_i$, the type of the corresponding parameter. 
In rule (T\_VAPPT), corresponding to the case where a variable bound to an anonymous function is applied, the function type is retrieved from $\Gamma$.

\subsection{Gradual Type System}\label{sec:gradual}


For the gradual type system we extend the types with the type \texttt{any} and the subtyping relation with the rule $\texttt{any} <: \texttt{any}$. 

In Figure~\ref{fig:typecheckinguntyped} we introduce the typing rules that consider gradualization. 
In the static type system we assumed that all function definitions have an associated type specification. In the gradual type system, however, we allow functions without type specifications. Those function definitions are considered untyped and are simply accepted by rule (CH\_FDEFU) without performing any type check. 

Rule (CH\_FDEFT) is a modified version of the one presented in Figure~\ref{fig:typechecking} in which we consider aspects of gradual typing in the function definition.
Those aspects are reflected in the premises $t' \lll t$ and $t'_i  \lll t_i$ of the rule. They state that the types of the function body and the patterns may be \emph{more precise} than the types specified in the function signature.

The \emph{precision relation}~\cite{garcia2013}%
\footnote{Also known as \emph{naive subtyping}~\cite{WF09}, \emph{less or equally informative}~\cite{SV08}, or \emph{materialization}~\cite{CLP+19} in the context of gradual typing.}, shown in Figure~\ref{fig:spec}, essentially says that the types $u$ and $t$ are related $u \lll t$ when $u$ results from changing in $t$ some occurrences of the type $\texttt{any}$ by other types. For instance, $\texttt{float} \lll \texttt{any}$, $[\texttt{bool}] \lll [\texttt{any}]$, $\{\texttt{any},\texttt{integer}\} \lll \{\texttt{any},\texttt{any}\}$.

Then, by rule (CH\_FDEFT), the following definition is correct, since $\texttt{integer} \lll \texttt{any}$.
\begin{lstlisting}
@spec func(integer) :: any
def func(x) do x end
\end{lstlisting}

\begin{figure}

\begin{prooftree} 
\AxiomC{$(\rho.f\!\_name, n) \not\in \Delta$}
\RightLabel{(CH\_FDEFU)}
\UnaryInfC{$\Delta; \rho \vdash^{ch} \texttt{def} \ f\!\_name (p_{1},...,p_{n}) \ \texttt{do} \ e \ \texttt{end}$}
\end{prooftree}

\begin{prooftree} 
\AxiomC{$\Delta(\rho.f\!\_name, n) = t_{1},...,t_{n} \rightarrow t $}
\noLine
\UnaryInfC{$\varnothing;\varnothing \vdash^{tp} p_{1} : t'_{1} \Rightarrow \Gamma^1  \quad ... \quad   \varnothing;\Gamma^{n-1} \vdash^{tp} p_{n} : t'_{n} \Rightarrow \Gamma^n \quad t'_i  \lll t_i $}
\noLine
\UnaryInfC{$\Delta; \Gamma^n; \rho \vdash^t e : t' \Rightarrow \Gamma' \quad t'  \lll t$}
\RightLabel{(CH\_FDEFT)}
\UnaryInfC{$\Delta; \rho \vdash^{ch} \texttt{def} \ f\!\_name (p_{1},...,p_{n}) \ \texttt{do} \ e \ \texttt{end}$}
\end{prooftree}

\hrule
\caption{Type Checking Rules for Gradual Types}
\label{fig:typecheckinguntyped}
\end{figure}

Precision is a reflexive relation (P\_REFL); rule (P\_ANY) states that all types are more precise than $\texttt{any}$.
The rest of the rules are structural; observe that, in contrast to subtyping, precision is covariant in all type positions. 

\begin{figure}

\begin{tabular}{c c}
\begin{minipage}{.35\columnwidth}
\begin{prooftree} 
\AxiomC{ }
\RightLabel{(S\_REFL)}
\UnaryInfC{$t \lll t$}
\end{prooftree}
\end{minipage}
&
\begin{minipage}{.42\columnwidth}
\begin{prooftree} 
\AxiomC{ }
\RightLabel{(P\_ANY)}
\UnaryInfC{$t \lll \texttt{any}$}
\end{prooftree}
\end{minipage}
\end{tabular}

\begin{prooftree} 
\AxiomC{$ t_{1} \lll u_{1} \ \ \cdots \ \ t_{n} \lll u_{n} \quad\quad t \lll u$}
\RightLabel{(P\_FUN)}
\UnaryInfC{$(t_{1},...,t_{n}) \rightarrow t \lll (u_{1},...,u_{n}) \rightarrow u$}
\end{prooftree}

\begin{tabular}{c c}
\begin{minipage}{.33\columnwidth}
\begin{prooftree} 
\AxiomC{$t \lll u$}
\RightLabel{(P\_LIST)}
\UnaryInfC{$[t] \lll [u]$}
\end{prooftree}
\end{minipage}
&
\begin{minipage}{.60\columnwidth}
\begin{prooftree} 
\AxiomC{$t_{i} \lll u_{i} \quad \forall i \in 1,..,n$}
\RightLabel{(P\_TUPLE)}
\UnaryInfC{$\{t_{1},..,t_{n}\} \lll \{u_{1},..., u_{n}\}$}
\end{prooftree}
\end{minipage}
\end{tabular}

\begin{prooftree} 
\AxiomC{$t_{i} \lll u_{i} \quad \forall i \in 1,..,n$}
\RightLabel{(P\_MAP)}
\UnaryInfC{$\%\{k_1 \Rightarrow t_{1},.., k_n \Rightarrow t_{n}\} \lll \%\{k_1 \Rightarrow u_{1},.., k_n \Rightarrow u_{n}\}$}
\end{prooftree}

\hrule
\caption{Precision Relation}
\label{fig:spec}
\end{figure}

\begin{figure}

\begin{prooftree} 
\AxiomC{$ \Delta; \Gamma; \rho \vdash^{t} e : t \Rightarrow \Gamma' \quad u \lll t$}
\RightLabel{(T\_DOWN)}
\UnaryInfC{$\Delta; \Gamma; \rho \vdash^{t} e : u \Rightarrow \Gamma' $}
\end{prooftree}

\hrule
\caption{Downcast of expressions}
\label{fig:expression:down}
\end{figure}

When a term with a gradual type $t$ occurs in a position where a term of a \emph{more precise} type $u$ is expected, our type system will perform a downcast, rule (T\_DOWN) in Figure~\ref{fig:expression:down}, provided that $u \lll t$. This rule together with the precision relation is the core of our approach to gradual typing. It follows the proposal of \cite{castagna} of using an antisymmetric relation and a subsumption-like rule for the downcast, instead of the more usual symmetric notion of \emph{consistency} \cite{ST06}.
 Notice that (T\_DOWN) introduces a sort of \emph{type unsafeness} as the downcast may lead to a runtime error caused by a type inconsistency, but this is the price to pay for dealing with code fragments that are untyped or gradually typed.

In Figure~\ref{fig:expression:functiongradual} we adapt the rules for function application to gradual typing. In our type system, untyped functions do not belong to $\Delta$. This is the case of both functions declared in the own program but without a type specification, and imported functions (i.e. legacy code).  Modeled by the new rules (T\_NAPPU) and (T\_LAPPU), for those functions we only verify that the arguments $e_i$ type check, but do not relate them to any expected parameter type. The return type for such functions is $\texttt{any}$. 

For instance, consider the expression
\texttt{2 + Main.fact(9)}, where function \texttt{Main.fact} is the presented in Section~\ref{sec:elixir}. When we apply the typing rules to this expression the rule (T\_ARITH) requires the type of the second argument to be $\texttt{integer}$. This is then required to the type of the function call, and that type is obtained by performing a downcast:

\begin{prooftree} 
\def\defaultHypSeparation{\hskip .1in}
\def\ScoreOverhang{1pt}
\AxiomC{$\texttt{Main.fact} \not\in \varnothing \quad ...$}
\RightLabel{\scriptsize{(T\_NAPPU)}}
\UnaryInfC{$\varnothing;\varnothing;\epsilon \vdash^{t} \texttt{Main.fact(9)} : \texttt{any} \Rightarrow \varnothing $}
\AxiomC{$\texttt{integer} \lll \texttt{any}$}
\RightLabel{\scriptsize{(T\_DOWN)}}
\BinaryInfC{\quad \quad \quad $\varnothing;\varnothing;\epsilon \vdash^{t} \texttt{Main.fact(9)} : \texttt{integer} \Rightarrow \varnothing $
}
\end{prooftree}

Rules (T\_NAPPT), (T\_LAPPT) and (T\_VAPPT), for functions with an available type specification, have been adapted to include the case where the arguments are of a \emph{more precise} type than the parameters.
To see an example, suppose we have a function \texttt{foo} in $\Delta$ with type $(\texttt{any}) \rightarrow \texttt{integer}$. Then the following is the derivation to type $\texttt{foo(9)}$, where the type of the argument ($\texttt{integer}$) is more precise than the type of the parameter ($\texttt{any}$).

\begin{prooftree} 
\def\defaultHypSeparation{\hskip .1in}
\def\ScoreOverhang{1pt}
\AxiomC{$ty(9) = \texttt{integer}$}
\RightLabel{\scriptsize{(T\_LIT)}}
\UnaryInfC{$\Delta;\Gamma;\epsilon \vdash^{t} \texttt{9} : \texttt{integer} \Rightarrow \Gamma 
$}
\AxiomC{$\Delta(\texttt{foo},1) = (\texttt{any}) \rightarrow \texttt{integer} $}
\noLine
\UnaryInfC{$\texttt{integer} \lll \texttt{any}$}
\RightLabel{\scriptsize{(T\_LAPPT)}}
\BinaryInfC{ $\Delta;\Gamma;\epsilon \vdash^{t} \texttt{foo(9)} : \texttt{integer} \Rightarrow \Gamma 
$}
\end{prooftree}

\begin{figure}

\begin{prooftree} 
\AxiomC{$\Delta(\rho_{2}.f\!\_name, n) = (t_{1},...,t_{n}) \rightarrow t $}
\noLine
\UnaryInfC{$\Delta; \Gamma; \rho_{1} \vdash^t e_{i}: t'_{i} \Rightarrow \Gamma^i \quad t'_i \lll t_i \quad \forall i \in 1,..,n $}
\RightLabel{(T\_NAPPT)}
\UnaryInfC{$\Delta; \Gamma; \rho_{1} \vdash^t \rho_{2}.f\!\_name \ (e_{1},..., e_{n}) : t \Rightarrow \Gamma^1 \dagger ... \dagger \Gamma^n$}
\end{prooftree}

\begin{prooftree} 
\def\ScoreOverhang{1pt}
\AxiomC{$(\rho_{2}.f\!\_name, n) \not\in \Delta $}
\noLine
\UnaryInfC{$\Delta; \Gamma; \rho_{1} \vdash^t e_{i}: t_{i} \Rightarrow \Gamma^i \quad \forall i \in 1,..,n $}
\RightLabel{(T\_NAPPU)}
\UnaryInfC{$\Delta; \Gamma; \rho_{1} \vdash^t \rho_{2}.f\!\_name \ (e_{1},..., e_{n}) : \texttt{any} \Rightarrow \Gamma^1 \dagger ... \dagger \Gamma^n$}
\end{prooftree}

\begin{prooftree} 
\AxiomC{$\Delta(\rho.f\!\_name, n) = (t_{1},...,t_{n}) \rightarrow t $}
\noLine
\UnaryInfC{$\Delta; \Gamma; \rho \vdash^t e_{i}: t'_{i} \Rightarrow \Gamma^i \quad t'_i \lll t_i \quad \forall i \in 1,..,n $}
\RightLabel{(T\_LAPPT)}
\UnaryInfC{$\Delta; \Gamma; \rho \vdash^t f\!\_name \ (e_{1},..., e_{n}) : t \Rightarrow \Gamma^1 \dagger ... \dagger \Gamma^n$}
\end{prooftree}

\begin{prooftree} 
\AxiomC{$(\rho.f\!\_name, n) \not\in \Delta $}
\noLine
\UnaryInfC{$\Delta; \Gamma; \rho \vdash^t e_{i}: t_{i} \Rightarrow \Gamma^i \quad \forall i \in 1,..,n $}
\RightLabel{(T\_LAPPU)}
\UnaryInfC{$\Delta; \Gamma; \rho \vdash^t f\!\_name \ (e_{1},..., e_{n}) : \texttt{any} \Rightarrow \Gamma^1 \dagger ... \dagger \Gamma^n$}
\end{prooftree}

\begin{prooftree} 
\AxiomC{$\Gamma(x) = (t_{1},...,t_{n}) \rightarrow t $}
\noLine
\UnaryInfC{$\Delta; \Gamma; \rho \vdash^t e_{i}: t'_{i} \Rightarrow \Gamma^i \quad t'_i \lll t_i \quad \forall i \in 1,..,n $}
\RightLabel{(T\_VAPPT)}
\UnaryInfC{$\Delta; \Gamma; \rho \vdash^t x \texttt{.} (e_{1},..., e_{n}) : t \Rightarrow \Gamma^1 \dagger ... \dagger \Gamma^n$}
\end{prooftree}

\hrule
\caption{Typing of Function Calls for Gradual Types}
\label{fig:expression:functiongradual}
\end{figure}


\section{Conclusions and Future Work}\label{sec:conclusions}

We have introduced a gradual type system \marcos{with subtyping} for a fragment of Elixir, which makes a compromise between static typing safety and dynamic typing flexibility. A prototype implementation of the type system is available at \cite{typelixir}.

Most approaches to gradual typing are based on a symmetric relation called \emph{consistency}~\cite{ST06} \marcos{and a notion of \emph{compatible subtyping}~\cite{ST07}, that puts together subtyping and consistency}. This is the case of \emph{Gradualizer}\cite{gradualizer}, a gradual type system for Erlang that uses the existing Erlang type specs
and provides support for parametric polymorphism and union types.  
\marcos{Another previous approach \cite{Tha89}, is based on subtyping, subsumption and down-casts; using the top type as the unknown type. Our type system is also based on subtyping, but we separate the top and unknown types. We follow the approach proposed by \cite{castagna}, where the subtyping and precision\footnote{In their case called \emph{materialization} and defined in the inverse direction.} relations have respective subsumption-like rules.}
\marcos{It is well known that the presence of a subsumption rule in a type system introduces problems in the implementation of type-checkers, because the rules are not syntax-directed. However, there exist also well known techniques to transform such systems to so-called algorithmic subtyping.}

\marcos{Most gradual type systems translate the code to a cast-calculus, in order to perform dynamic checks (only) to the gradual fragments. We do not perform this step, since one of our main goals has been not to interfere with Elixir's compiler code generation phase, which translates to the dynamically typed BEAM language. However, a possible line of future work is to define a translation to a cast-calculus and investigate how this can be used by Elixir's compiler to improve the generated code.}

To be completely sure about our formalization, we still need to prove properties of our type system, like soundness or subject reduction. For that we first need to formalize the operational semantics of the language. \marcos{We have implemented a formalization of an important part of the type system in Idris and we are now in the process of proving some of the desired properties\footnote{\url{https://gitlab.fing.edu.uy/elixir/gradualelixir/}}}. 
In a system like the one we presented progress does not hold because some checks (originated by downcasts) are delegated to runtime, and it is possible that they fail producing that reduction get stuck. However, we think that progress would hold in our system for fully-statically-checked well-typed programs (i.e. programs that do not require downcasts). 
Another interesting property to prove is gradual guarantee \cite{siek2015}, which states that removing the type specifications from a gradually typed program always produces a program that is still well typed.

In the actual setting, the definition of a pseudo-polymorphic function implies to loose the typing properties of the function. As future work, we plan to introduce proper parametric polymorphism to our type system.
Another line of future development is to enlarge the number of language features covered by the type system. 

Finally, we also plan to follow the methodology proposed by Cimini and Siek \cite{cimini}, which uses a tool called \emph{gradualizer} to generate a gradual type system starting from a static one. Our idea is to generate a new version of the gradual type system using that methodology and compare the resulting type system with ours.


 \bibliographystyle{elsarticle-num} 
 \bibliography{biblio}





\end{document}